\documentstyle[preprint,aps,tighten]{revtex}
\newcommand{\ds}{\displaystyle}
\begin{document}
\draft
\title{Elementary Quantum Mechanics in a  Space-Time Lattice}
\author{Manjit Bhatia$^1$}
\address{Department of Mathematics, University of Maryland, College Park, MD 20742, U.S.A. \\and\\ Professor Emeritus, Bowie State University, Bowie, MD 20715, U.S.A }
\author{P. Narayana Swamy$^2$}
\address{Professor Emeritus, Department of Physics, Southern Illinois University,\\
Edwardsville, IL 62026, U.S.A.}
\maketitle
\begin{abstract}
Studies of quantum fields and gravity suggest the existence of a minimal length, such as Planck length \cite{Floratos,Kempf} and we ask how this may modify the results in elementary quantum mechanics (QM) problems familiar to us \cite{Gasiorowicz}. In this paper, we examine the problem from a semi-classical phenomenological point of view. We assume a lattice model for (1+1)-space-time, which consists of a grid of $\lambda_0 \times \tau_0 $ rectangles, where $\lambda_0$, the lattice parameter, is a fundamental length (say Planck length) and, we take $\tau_0$ to be equal to $\lambda_0/c$. We obtain a complete solution of a simple problem from elementary non-relativistic quantum mechanics where the usual continuum (1+1)-space-time is supplanted by the proposed space-time lattice. It is our goal to bring out the $\lambda_0$-dependence of results for eigenvalues, eigenfunctions, uncertainties in position and in momentum of the particle,  in the problem under discussion. We address  the familiar problem, ``particle in a box", described by the corresponding Schrodinger difference equation, the solution of which yields the $q$-eigenfunctions and $q$-eigenvalues of the energy operator as a function of $\lambda_0 $.  The $q$-eigenfunctions are normalizable and we determine the eigenvalues. We then compute the uncertainties in position and momentum, $\Delta x, \Delta p$ for the box problem and study the consequent modification of Heisenberg uncertainty relation due to the assumption of space-time lattice, in contrast to modifications suggested by other investigations \cite{Floratos}. We show that as the lattice parameter $ \lambda_0 $ approaches zero, the results obtained under lattice assumption converge to the corresponding results obtained under the usual continuum quantum mechanics.

\end{abstract}
 \noindent ($1$)  Electronic address: mbhatia@math.umd.edu\\
 ($2$) Electronic address: pswamy@siue.edu \\

\section{Introduction}

It is widely believed \cite{Calmet} that a minimum length, of the order of the Planck length, results from combining quantum mechanics and classical general relativity. The key ingredients used to reach this conclusion are the  uncertainty principle of quantum mechanics (QM) together with general relativity \cite{Kempf}. In recent years, considerable effort has been expended in exploring implications of the existence of a fundamental length, such as Planck length, in non-relativistic QM \cite{Maziashvili,Akhoury}.

In this paper we assume the existence of a fundamental length, $\lambda_0$, as given regardless of its origin. We assume that $\lambda_0$ is the smallest measurable length in space dimension and  $\tau_0=\lambda_0/c$ is the smallest measurable interval for the time dimension. This assumption
enables us to model (1+1) dimensional space-time as a lattice (grid).  We then propose to compare the solutions of Schrodinger equation for an elementary QM problem under two contrasting assumptions regarding space-time namely:\\\\
(1) the usual (1+1) dimensional continuum for space-time.\\
(2) a (1+1) dimensional space-time  lattice (grid) of $\lambda_0 \times \tau_0$ rectangles.\\

The lattice described above is our (rather simplistic, if not naive) model for `quantized' space-time. Under this model, we are interested in the dependence of eigenvalues, eigenfunctions, their orthogonality and normalization on the lattice parameter, $\lambda_0$. We also investigate the $\lambda_0$-dependence of average uncertainties in position and in momentum for the problem at hand. This enables us to check what bearing, if any, $\lambda_0$ has on the Heisenberg uncertainty relation in our model.

In elementary non-relativistic quantum mechanics, under the assumption of space-time continuum,  starting from the Hamiltonian of a one-dimensional particle in a potential $V(x)$, the time-dependent Schrodinger equation for the wave function $\psi(x,t)$ can be written as
\begin{equation}
\frac{-\hbar^2}{2m} \frac{\partial^2 \psi(x,t)}{\partial x^2}+ V(x)\psi(x,t) =i \hbar\frac{\partial \psi(x,t)}{\partial t}  \label{SchEqn}
\end{equation}
With $\psi(x,t)=T(t)u(x)$, (\ref{SchEqn}) splits into two ordinary differential equations in the usual manner.
\begin{equation}
\frac{dT(t)}{dt}+i\omega T(t) =0  \label{SEqnt}\, ,
\end{equation}
\begin{equation}
\frac{d^2u(x)}{d^2 x}+k^2 u(x)=0  \label{SEqnx} \, ,
\end{equation}
where $\omega ={\displaystyle \frac{E}{\hbar}}$ and $k^2={\displaystyle \frac{2m[E-V(x)]}{\hbar^2}}$ \label{omgk2}.\\

The general solution of Schrodinger equation (\ref{SchEqn}), can be obtained once $V(x)$ and the relevant boundary conditions are specified.\\

The rest of the paper is organized as follows. In Section II, we summarize the solution of the well-known `particle in a box' problem under the usual assumption of space-time continuum. This is one of the standard  problems discussed in introductory quantum mechanics textbooks ( for example, \cite{Gasiorowicz}). In addition to eigenvalues and eigenfunctions for the problem in the  continuum case, we also include the expressions for uncertainties in position and momentum, $\Delta x$ and $\Delta p$, respectively. This is done for subsequent comparison of these quantities with the corresponding results for solutions obtained under the assumption of space-time lattice (hereafter called just lattice). We derive the eigenfunctions and eigenvalues under the lattice assumption in Section III (denoting them as {\it q}-eigenvalues and {\it q}-eigenfunctions). We show that as the lattice parameter $\lambda_0$ $\rightarrow 0$, the {\it q}-eigenvalues and {\it q}-eigenfunctions approach the corresponding continuum eigenfunctions and eigenvalues. We devote Section IV to the investigation of  orthogonality and normalization of {\it q}-\nolinebreak[4]eigenfunctions. The uncertainties $\Delta x$ and $\Delta p$  under the lattice assumption are computed.  Again, we observe that as we proceed to the limit $\lambda_0 \rightarrow 0 $, all the  results for $\Delta x$ and $\Delta p$ obtained under the lattice assumption converge to the corresponding continuum results. A summary of our results and some remarks about possible future work are  given in Section V.

\section {Particle in a box - Solution with Space-Time Continuum}

In this section we state the problem of `particle in a box' in a notation similar to that of  \cite{Gasiorowicz} and summarize the standard solution. The size of the box is denoted by $L$.
The potential function $V(x)$ is defined by:

\vspace{.1in}

\hspace*{2.0in}$V(x)=\left\{ \begin{array}{lll}
 \infty & x \leq 0   \\
  0 & 0 < x < L   \\
 \infty & x \geq L    \label{pot}
\end{array}  \right.$\\

The nominal domain of the function $u(x)$ is the set of all reals, ${\bf R} $, but the interesting part of the domain is the interval  \(I = \{x \in R \:|\; 0 \leq x \leq L \} \).
The function $u(x)$ must satisfy the boundary conditions:
\begin{equation}
u(x) = \cases{0  &if {\it x} $\leq 0$ \cr  0 & if {\it x} $\geq$ L \cr}
   \label{bcond}\, .
\end{equation}

\subsection{Eigenvalues and Eigenfunctions with Space-time Continuum}
 The solution of (\ref{SEqnt}) is given by:
\begin{equation}
T(t)=T_0 e^{i \omega t}  \, . \label{Gsolt}
\end{equation}
Setting  ${\displaystyle k^2 = \frac{2m E}{\hbar^2}}$, $(E>0)$, the time independent Schrodinger equation (\ref{SEqnx}) simplifies to the equation
\begin{equation}
\frac{ d^2u(x)}{dx^2}+ k^2 u(x) =0, \;\;\;\;\;\; 0 < x < L   \label{SEqn1}
\end{equation}

The general solution of  (\ref{SEqn1}) in the interval $0 < x < L $  subject to boundary conditions  (\ref{bcond}) under the assumption of a space-time continuum is discussed in elementary quantum mechanics textbooks
( for example \cite{Gasiorowicz}), and can be written as:
\begin{eqnarray}
        u_n(x)&=&   B \left[ e^{ikx}  - e^{-ikx} \right] \label{ceigenf}\, ,
             \end{eqnarray}
    \begin{eqnarray}
\emph{i.e.,} \;\; u_n(x)&=& 2 i B \; \sin(kx) \label{ceigenf2}
    \end{eqnarray}
where eigenvalues for $k$ are given by ${\displaystyle k = \frac{\pi n}{L}}$, $n = 1, 2, 3, \ldots$.\\
Substituting for $k$  and setting $A=2iB$, we can write the eigenfunctions in the usual form
\begin{eqnarray}
 u_n(x)&=& A \sin(\frac{n\pi x}{L}) \label{ceigenf2}
    \end{eqnarray}
and exhibit the energy eigenvalues as:
\begin{equation}
E_n=  \ds \frac{\hbar^2 k^2 }{2m}=  \frac{\hbar^2 \pi^2 n^2}{2mL^2}, \;\; n = 1, 2, 3, \ldots \label{ceigenv}
\end{equation}
For $n \ne n'$, the eigenfunctions   $u_n(x)$ and  $u_{n'}(x)$ are orthogonal and the normalization constant $A$ is determined to be
\begin{equation}
|A|=\sqrt{\frac{2}{L}}. \label{normal}
\end{equation}
The uncertainties $ \Delta x $ and $ \Delta p $ can also be easily computed. We have:
\begin{eqnarray}
\Delta x &=& \sqrt{|< x^2 > - < x >^2 |} = \sqrt{\frac{ n^2 \pi^2 - 6}{3}} \cdot \frac{L}{2 n \pi}  \label{cdelx}  \\
\Delta p &=& \sqrt{|< p^2 > - < p >^2|\;}  = \frac{n \pi \hbar}{L}   \label{cdelp} \\
\Delta x \; \Delta p &=& \sqrt{\frac{ n^2 \pi^2 - 6}{3}} \cdot ( \hbar /2)  \geq  \frac{\hbar}{2} \;\;
( {\rm for \;\; all \;\; } n \geq 1 ) \, .  \label{ccur}
\end{eqnarray}
To no one's surprise, (\ref{ccur}) verifies the validity of the Heisenberg uncertainty relation for the particle confined in a box.

\section {Particle in a box - Solution with Space-Time Lattice}
Next we proceed to obtain the solution of (\ref{SEqn1}), with boundary conditions (\ref{bcond}) under the assumption of quantized space-time.

We use the word ``quantized'' as synonymous with discreteized. Thus our quantized $x$-domain would simply be the discreteized  real line given by $R_{\lambda_0} = \{j \lambda_0 \; |\; j \in {\bf Z} \}$, where $\lambda_0$ is the lattice parameter introduced earlier. (The lattice parameter $\lambda_0$ could be a ``fundamental'' length, such as the Planck length \cite{Baez,Planck,Adler}) and $j$ is an arbitrary integer. For simplicity, we express the quantized time domain in terms of $\tau_0 = \lambda_0/c$, $c$ being the velocity of light. Thus $R_{\tau_0} = \{j_t \tau_0 \;|\; j_t \in \mathbf{Z} \}$.  Note that for any integer $j$, the quantity $j \lambda_0$ corresponds to a particular point on the real line $\mathbf{R}$. Similarly $j_t \tau_0$ gives us a specific value for time $t$ for any integer $j_t$. The continuum limit would  obtain if  $j \rightarrow \infty.
  $ and
$\lambda_0 \rightarrow 0$ then $j \lambda_0 \rightarrow x$, where $x$ is a real number, $ 0 \leq x \leq L$.
Similarly for $j_t$ and $j_t \tau_0$. \\
\subsection{Eigenvalues and Eigenfunctions with Space-time Lattice}
The time-dependence of functions T(t) in the space-time setting can be presented as a simple (central\footnote{See for example \cite{Hildebrand}. It turns out the central difference representation for the differential operators leads to a hermitian representation for all relevant physical quantities. })  difference equation, (we use superscript $q$ to denote results  under our space-time lattice model)  thus:
\begin{equation}
\frac{T^q((j_t+1)\tau_0) - T^q((j_t - 1)\tau_0)}{2 \tau_0} +i\omega T^q(j_t\tau_0)= 0 \label{qDift1}
\end{equation}
or
\begin{equation}
T^q((j_t+1)\tau_0) - T^q((j_t - 1)\tau_0) +2 i \tau_0 \omega T^q(j_t\tau_0)= 0 \label{qDift}
\end{equation}

The solution of the above difference equation can be readily seen to be
\begin{equation}
T^q(j_t\tau_0) = T^q_0 e^{ i j_t \theta }, \;  \theta = \arctan\left(\frac{-w\tau_0}{\sqrt{1 - w^2 \tau_0^2}} \right) \label{qGsoltl}
\end{equation}
where we have set $T^q_0= T^q(0)$, the value of the function at time zero, in agreement with (\ref{Gsolt}). We note that for small $\lambda_0$ (and therefore small $\tau_0$, the above solution reduces to:
 \begin{equation}
  T^q(j_t\tau_0) = T^q_0 e^{- i w j_t \tau_0 }  \label{qGsoltsmpl}
 \end{equation}
We will not have much occasion to discuss  the time dependence of eigenfunctions in this paper. We assume that the length $L$ of the box is an integer multiple of \nolinebreak[4] $\lambda_0$ \nolinebreak[4]. Thus
\begin{equation}
 L=J_0 \lambda_0            \label{jLdef}
\end{equation}
where $J_0$ is a positive integer. For the quantized real line we can restate the boundary conditions (\ref{bcond}) as
\begin{equation}
u^q(j\lambda_0)  = \cases {0 \; & if $ j \leq 0$   \cr
  0 \; & if $j \geq J_0,$ \cr}\;\;    \label{qbcond0}
\end{equation}
where $j \in {\bf Z}$.

Next, we replace the `second derivative' on the left hand side  of (\ref{SEqn1}) by the `second (central) difference' with the prescription (see, for example, \cite{Hildebrand}):
\begin{equation}\label{Difdp}
  \frac{ d^2u(x)}{dx^2}  \longrightarrow  \frac{u^q((j+2)\lambda_0)-2u^q(j\lambda_0)+u^q((j -2)\lambda_0)}{4 \lambda_0^2}\, .
\end{equation}
It is significant to point out that we are employing the ``central difference". The various differences are defined in page 7 of Hildebrand \cite {Hildebrand}. Carrying out the replacement in (\ref{SEqn1}) and multiplying
through by $(4\lambda_0^2 )$, we have:
\begin{equation}
u^q((j+2)\lambda_0)-2(1-2 k^2 \lambda_0^2) u^q(j \lambda_0)+ u^q((j-2)\lambda_0) = 0  \label{DSEqn5}
\end{equation}
where \begin{equation}
k^2 = \frac{2 m E}{\hbar^2}   \label{defk2}
\end{equation}

The above equation is a difference equation of  second order with constant coefficients and can be solved by standard methods. The main steps are outlined below. We use the ansatz $ u^q(j \lambda_0) = a^j $. Then the equation (\ref{DSEqn5}) becomes
\begin{equation}
a^{j+2} -2(1-2 k^2 \lambda_0^2) a^j + a^{j-2} = 0  \label{SDSEqn5a}
\end{equation}
from which we infer that
\begin{equation}
 a = e^{i \theta /2}, \label{SDSEqn5ee}
\end{equation}
where $\theta $ is given by
\begin{equation}
 \tan{\theta} = \pm \frac{ 2 k \sqrt{1 - k^2}}{1 - 2 k^2} \label{SDSEqn5f}\, .
 \end{equation}

The general solution of (\ref{DSEqn5}) can thus be exhibited as:
\begin{equation}
u^q(j \lambda_0) = A e^{i j \theta/2}  +  B e^{- i j \theta/2}, \label{LatSol1a}
\end{equation}
where $\theta$ is given by (\ref{SDSEqn5f}. Imposing the boundary conditions (\ref{jLdef}), we get
\begin{eqnarray}
A+B=0                                                 \label{letbc1a} \\
Ae^{i J_0 \theta/2} + B e^{-i J_0 \theta/2} = 0      \label{letbc1b}
\end{eqnarray}
which lead to the eigenvalue equation:
\begin{equation}
e^{i J_0 \theta} =1  \label{letegnv1a}
\end{equation}
with the solution
\begin{equation}
\theta=\frac{2n\pi \lambda_0}{J_0}, \;\;\; {\rm with} \; n= 1, 2, 3, \ldots
\end{equation}
We thus obtain
\begin{equation}
 \frac{ 2 k \lambda_0 \sqrt{1 - k^2 \lambda_0^2}}{1 - 2 k^2 \lambda_0^2} = \tan{\frac{2 n \pi \lambda_0} {L }} \, .    \label{letegnv1d}
\end{equation}
We can solve (\ref{letegnv1d}) to obtain a closed expression for $k^2$ and consequently $E_n$ as
shown below. For convenience, we define
\begin{eqnarray}
y=k^2 \lambda_0^2  \;\;\;\; {\rm and } \;\;\;\; t = \tan{\left( \frac{2 n \pi \lambda_0 } {L } \right) }. \label{eigenk2def}
\end{eqnarray}
 The equation (\ref{letegnv1d}) then becomes:
\begin{equation}
\frac{2 \sqrt{y} \sqrt{1-y} }{1 - 2 y} = t                  \label{solndtls1} \\
\end{equation}
and we obtain
\begin{equation}
  y = \frac{1 + t^2 -\sqrt{1 + t^2}}{2( 1 + t^2)}\, . \label{solndtls6}
\end{equation}
Substituting for $y$ and $t$ in (\ref{solndtls6}), and observing that
$$ 1 +t^2 = 1 + \tan^2{\left( \frac{2 n \pi \lambda_0} {L } \right) }  \; = \; \sec^2{\left(\frac{2 n \pi \lambda_0} {L } \right) } $$
gives us the following equation. \\
\begin{eqnarray}
  k^2 = \frac{1}{2 \lambda_0^2}\left (1 - \cos{(\frac{2n \pi \lambda_0}{L})} \right )     \label{eigenk2} \\
  \Rightarrow E^q_n = \frac{\hbar^2}{4 m \lambda_0^2}\left (1 - \cos{(\frac{2n \pi \lambda_0}{L})} \right )\label{eigenEn}
\end{eqnarray}
Thus, as in the continuum case, the variable $k$ is  discretized by the requirement that
the solution of (\ref{DSEqn5}) satisfy the boundary conditions ( \ref{letbc1a}, \ref{letbc1b}). Expanding the right hand side  of (\ref{eigenEn}) in powers of $\lambda_0$, we obtain
\begin{eqnarray}
E^q_n &=& \frac{\hbar^2}{4 m \lambda_0^2}\left(1 - \cos{\left(\frac{2 n \pi \lambda_0}{L}\right)} \right) \label{eigenEnsrs1}\\
&=& \frac{\hbar^2}{4 m } \left(\frac{2 n^2 \pi^2 }{L^2} -
\frac{2 n^4 \pi^4}{3 L^4}\lambda_0^2 + O(\lambda_0^4) \right) \label{eigenEnsrs2}\\
 &=& \frac{\pi^2 \hbar^2 n^2}{2 m L^2} - \frac{\pi^4 \hbar^2 n^4}{6 m L^4} \lambda_0^2
+ O(\lambda_0^4),  \;\;\; \;\; n = 1, 2, 3, \ldots \label{eigenEnsrs2}
\end{eqnarray}

The second term in the previous equation is the leading correction to $E_n$, due to the assumed fundamental length $\lambda_0$, which turns out to be $O(\lambda_0^2)$. We note that
our results for correction to eigenvalues of `particle in box' problem are consistent with the corrections of hydrogen atom spectrum (also $O(\lambda_0^2)$) obtained by  Maziashvili et al \cite{Maziashvili} (treating it as a quantum-gravitational effect) and by Akhoury et al \cite{Akhoury} (using minimum-length deformed QM) \footnote{See \cite{Maziashvili} for a discussion of sign difference between the results for corrections obtained by  \cite{Maziashvili} and \cite{Akhoury}}.\\

In the limit as $ \lambda_0 \rightarrow 0 $, we have  $ E^q_n \rightarrow E_n $, (see (\ref{ceigenv})) a result which serves as a check on our calculation.

Since $k^2$ is nonnegative, $\sqrt{k^2}=\pm k$ is a real number. With $k$ determined by (\ref{eigenk2}), the eigenfunctions $u^q_n(j\lambda_0)$ are given by
\begin{equation}
         u^q_n(j\lambda_0)=   A^q  \sin{(n \pi j \lambda_0 /L)}   \label{qeigenf2}
    \end{equation}
To determine the normalization factor, $A^q$, and to check the orthogonality of the eigenfunctions $u^q_n(j\lambda_0)$,  we  have to compute the following sum sum. This is somewhat tedious to do by hand but use of \emph{Mathematica} (or some other similar mathematical software) makes the analysis quite simple. We obtain:
\begin{eqnarray}
 S_{nn'}=& \sum_{j=0}^{J_0} u^{q*}_{n'}(j\lambda_0) u^{q}_n(j\lambda_0)   \label{ortho1} \\
        =& |A^q|^2 \sum_{j=0}^{J_0} \left[ \sin{(n' \pi j \lambda_0 /L)} \; \sin{(n \pi j \lambda_0 /L)}  \right]   \label{ortho2} \\
        =& |A^q|^2 \left(\frac{L}{2} \right) \delta_{nn'}    \label{ortho3}
 \end{eqnarray}
 where $\delta_{nn'}$ is the usual Kronekar delta symbol. It is then clear that the $q$-eigenfunctions $u^q_n(j\lambda_0)$ are indeed mutually orthogonal for $n \ne n'$ and that the normalization factor $$ |A^q| = \sqrt{\frac{2}{L}}$$
which turns out to be the same as in the continuum case.

\section{Computation of Uncertainties $\Delta x$ and $\Delta p$ with Space-Time Lattice}
In the present section we calculate the uncertainties $\Delta x$ and $\Delta p$ with $q$-eigenfunctions in space-time lattice as given in (\ref{qeigenf2}).
As in the continuum case, we compute  $< p >_n, \;\;\;< p^2 >_n, \;\;\ < x >_n$ , and $< x^2 >_n  $, and using these we obtain $\Delta x \;$ and $\; \Delta p $ as usual.\\\\
{\bf Computation of $\Delta p$ }:\\

Recall that the operator $p$ in central derivative representation takes the form
\begin{eqnarray}
p u^q_n(j \lambda) \rightarrow&   -i \hbar \; \nabla u^q_n(j \lambda_0)    \label{delp1} \\
\rightarrow& - i \hbar \;\frac{u^q_n\left((j+1) \lambda_0 \right) - u^q_n\left((j-1) \lambda_0 \right)}{2 \lambda_0} \label{delp2}
\end{eqnarray}
Using ( \ref{qeigenf2}), we obtain
\begin{eqnarray}
< p>_n &=& \sum_{j=0}^{J_0} u^{q*}_{n}(j\lambda_0) p u^{q}_n(j\lambda_0) \label{delp3}\\
& = &\frac{|A^q|^2}{ 2 \lambda_0} \left[ \sum_{j=1}^{J_0}  \; \sin{(n \pi (j-1) \lambda_0 /L)}\sin{(n \pi j \lambda_0 /L)}\right.\\\nonumber
&-& \left.\sum_{j=1}^{J_0} \sin{(n \pi j \lambda_0 /L)}\sin{(n \pi (j-1) \lambda_0 /L)}   \right] \label{delp5}\\
& =& 0    \label{delp5a}
\end{eqnarray}
where we added and subtracted some terms to the two sums in (\ref{delp5}), which vanish because of boundary conditions (\ref{qbcond0}).
Thus we see that  $< p>_n = 0 $.\\

We can compute $< p^2>_n $ by evaluating the sum directly. We can also get the answer more readily by observing that
 \begin{equation}
 < p^2>_n = < 2 m E >_n = \hbar^2 k^2 =  \frac{\hbar^2}{2 \lambda_0^2} [1- \cos (\frac{2 n \pi \lambda_0}{L}) ]
 \end{equation}
Thus
 \begin{eqnarray}
 (\Delta p)^2= & < p^2>_n -< p>_n^2   \label{delpsqd0}  \\
  = & \frac{n^2 \pi^2 \hbar^2}{L^2} - \frac{n^4 \pi^4 \hbar^2 }{3 L^4} \lambda_0^2 + O(\lambda_0^4)    \label{delpsqd2}
 \end{eqnarray}
 and therefore
 \begin{eqnarray}
 \Delta p = & \left(\frac{\hbar}{\lambda_0}\right) \sin{(\frac{n \pi \lambda_0}{L})} \label{delp0} \\
  = & \frac{\pi  n \hbar}{L} - \frac{\pi^3 n^3 \hbar}{6 L^3} \lambda_0^2 + O(\lambda_0^4)   \label{delp1}
  \end{eqnarray}
\\{\bf Computation of $< x >_n $: }

\vspace{.1in}

Since
 \begin{eqnarray}
 u^{q*}_{n}(j\lambda_0)x u^q_n(j \lambda) \rightarrow&   u^{q*}_{n}(j\lambda_0) j \lambda_0 u^q_n(j \lambda_0)    \label{delx1a}
\end{eqnarray}
using (\ref{qeigenf2}), and recalling that
\begin{eqnarray}
a = (n \pi  \lambda_0 /L).   \label{delx1b}
\end{eqnarray}
we obtain
\begin{eqnarray}
< x >_n\;=& \lambda_0^2\; \sum_{j=0}^{J_0} \; j\; u^{q*}_{n}(j\lambda_0)\; u^{q}_n(j\lambda_0)
\label{delx1c} \\
        =& |A^q|^2 \;\lambda_0^2 \;\sum_{j=0}^{J_0} j \sin^2{(a j)}    \label{delx1d}
\end{eqnarray}\\
The sum $\sum_{j=0}^{J_0} j \sin^2{(a j)}$ can be evaluated by standard techniques and we obtain:
\begin{eqnarray}
\sum_{j=0}^{J_0} j \; \sin^2{(a j)} = \frac{L^2}{4}  \label{sumj1}
\end{eqnarray}
where we have used the definition $J_0 = \frac{L}{\lambda_0} $ to simplify the final sum.
Substituting for $|A^q|^2$ in (\ref{delx1d}), we get:
\begin{eqnarray}
< x >_n\; =& \frac{L}{2}    \label{expctx1}
\end{eqnarray}
\\{\bf Computation of $<x^2>_n$: }

\vspace{.1in}

Now since
 \begin{eqnarray}
 u^{q*}_{n}(j\lambda_0)x^2 u^q_n(j \lambda) \rightarrow&   u^{q*}_{n}(j\lambda_0) (j \lambda)^2 u^q_n(j \lambda_0)    \label{delx2a}
\end{eqnarray}
using (\ref{qeigenf2}),
we obtain
\begin{eqnarray}
< x^2 >_n \;=& \lambda_0^3\; \sum_{j=0}^{J_0} \; j^2 \; u^{q*}_{n}(j\lambda_0)\; u^{q}_n(j\lambda_0)
\label{delx2c} \\
        =& |A^q|^2 \;\lambda_0^3 \;\sum_{j=0}^{J_0} j^2 \sin^2{(a j)}    \label{delx2d}
\end{eqnarray}\\
The sum $\sum_{j=0}^{J_0} j^2 \sin^2{(a j)}$ can be evaluated by standard techniques and we obtain:
\begin{eqnarray}
\sum_{j=0}^{J_0} j^2 \; \sin^2{(a j)} =& \frac{L^3}{6} + \frac{L \lambda_0^2}{12}
- \left(\frac{L \lambda_0^2}{4}\right) \csc^2{(\frac{\lambda_0 n \pi}{L})} \label{sumj2a} \\
\texttt{}=& \left( \frac{L^3}{6} - \frac{L^3}{4 n^2 \pi^2} \right) - \left( \frac{n^2 \pi^2}{60 L}\right)\lambda_0^4 -O( \lambda_0^6)  \label{sumj2b}
\end{eqnarray}
Substituting for $|A^q|^2$, and rearranging,  we obtain:
\begin{eqnarray}
< x^2 >_n\; =& \frac{L^2}{3} + \frac{\lambda_0^2}{6} - \left(\frac{\lambda_0^2}{2}\right) \csc^2{(\frac{\lambda_0 n \pi}{L})} \label{expctx2}
\end{eqnarray}
Using (\ref{expctx1}) and (\ref{expctx2}), we get:
\begin{eqnarray}
(\Delta x)^2 =& < x^2 >_n - (<x>_n)^2    \label{Deltaxsqd1} \\
=& \left(\frac{L^2}{12} - \frac{L^2}{2\; n^2 \; \pi^2 } \right) - \left(\frac{n^2 \; \pi^2}{30 \; L^2}\right) \lambda_0^4 + O(\lambda_0^6) \label{Deltaxsqd3}
\end{eqnarray}\\
which leads to
\begin{eqnarray}
\Delta x =& \left( \frac{1}{2 \; \sqrt{3} \;}\right) \; \sqrt { \;L^2 + \; 2 \; \lambda_0^2 - 6 \; \lambda_0^2 \; \csc^2{(\frac{\lambda_0 n \pi}{L})}} \;  \label{Deltax1} \\
=& \left(\sqrt{\frac{L^2}{12} - \frac{L^2}{2 \; n^2 \; \pi^2 }} \right) - \left(\frac{n^2 \; \pi^2}{60 \left( \; L^2 \; \sqrt{\frac{L^2}{12} - \frac{L^2}{2\; n^2 \; \pi^2 }}\right) } \right) \lambda_0^4 \; + \; O(\lambda_0^6) \label{Deltax2}
\end{eqnarray}
Using equations (\ref{Deltax1}) and (\ref{delp1}), we can write:
\begin{eqnarray}
\Delta x \Delta p =& \left( \frac{\hbar}{2 \; \sqrt{3} \; \lambda_0}\right) \left(\; \sqrt { \;L^2 + \; 2 \; \lambda_0^2 - 6 \; \lambda_0^2 \; \csc^2{(\frac{\lambda_0 n \pi}{L})}} \; \right)
\sin{(\frac{n \pi \lambda_0}{L})} \label{dxdp1} \\
=& \left[ \left( \sqrt{ \frac{n^2 \pi^2 -6} { 3 }} \right) - \left(
\frac{ n^2 \pi^2 \sqrt {\frac{n^2 \pi^2 -6}{ 3 }}}{6 L^2} \right)\lambda_0^2 \right]
\left( \frac{\hbar}{2} \right) + O(\lambda_0^4) \label{dxdp2}\\
=& \sqrt{\frac{n^2 \pi^2 -6} { 3 }}\left[ 1 - \left(\frac{n^2 \pi^2}{6 L^2} \right) \lambda_0^2 \right] \left(\frac{hbar}{2} \right) +O(\lambda_0^4) \label{dxdp3}
\end{eqnarray}
The first term in the square bracket in (\ref{dxdp2}) gives us the continuum result while the second term is the correction because of the assumed space-time lattice and is of  second order in the
lattice parameter $\lambda_0$.

\section{Particle in a box: Summary}

With our model of quantized space-time as  described in the foregoing, we have obtained the $q$-eigenvalues and $q$-eigenfunctions for the problem of a particle in a box as a function of the lattice parameter $\lambda_0$. We have shown that the $q$-eigenfunctions can be normalized and that the eigenfunctions are {\it not} mutually orthogonal. We find that the `degree of nonorthogonality' is  $O(\lambda_0^2)$. We have also computed the uncertainties $\Delta x $ and $ \Delta p $  for the problem. The modification of the Heisenberg uncertainty relation ( see (\ref{dxdp2}, \ref{dxdp3}) ) induced by the assumption of space-time lattice is explicitly exhibited up to second order in lattice parameter $\lambda_0$.\\

By examining the limit process as $\lambda_0 \rightarrow 0$, we can see that the $q$-eigenvalues  reduce to the ones obtained for the case of continuum domain. Same holds for the $q$-eigenfunctions as also for the expressions for $\Delta x, \Delta p$. This reflects the continuous nature of the results for eigenfunctions, eigenvalues and uncertainties  obtained for the space-time lattice, as a function of the lattice parameter $ \lambda_0 $.\\

 The first order modification of product $\Delta x \Delta p $ induced by the assumed space-time lattice is of the form $$\Delta x \Delta p = \alpha \left(1 + \beta \; \lambda_0^2 \right)\left(\frac{\hbar}{2}\right)$$ where the coefficients $\alpha$ and $\beta$ for the `particle in box' problem, $\beta$ can be read off from (\ref{dxdp3}) as
\begin{eqnarray}
 \alpha = \sqrt{\frac{n^2 \pi^2 -6} { 3 }}  \\ \label{alpha}
 \beta =- \left(
\frac{ n^2 \pi^2 }{6 L^2} \right) \label{beta}
 \end{eqnarray}\\
 The parameters  $\alpha$ and $\beta$, of course, depend on the box problem that we are investigating. However the form of correction of the product of uncertainties  above is similar to (but different from) the one given in \cite{Floratos} as well as other papers exploring consequences of minimum length in uncertainty relation (see, for example, \cite{Maziashvili} and \cite{Akhoury}). It is interesting to note that even a simplistic for space-time lattice and semi-classical approach to QM seems to reveal shadows of deeper theories.

\section{ Concluding Remarks}

Motivated by the possible existence of a fundamental length, such as Planck length, as a physical reality, we pose the question: how may a fundamental length modify the results in elementary quantum mechanics problems familiar to us? The investigation of the implications of space-time lattice from many different perspectives continues to draw considerable attention in literature (See for example \cite{Floratos}).  In this paper, we have explored the ``particle in a box" problem in some detail with a simple model for space time lattice and obtained corrections induced by the existence of a space-time lattice to the results for eigenvectors eigenvalues, average uncertainties given under the space-time continuum. We have exhibited how the Heisenberg uncertainty relation in space-time continuum compares with the corresponding result in space-time lattice.

For the ``particle in a box" problem with space-time lattice, we found that  the corrections to eigenvalues etc are of second order in the lattice parameter $\lambda_0$, and although the corrections are observable in principle, the rather small magnitude of Planck length, the lattice parameter $\lambda_0 \approx 10^{-35}$ renders it difficult if not impossible to measure at present. All the same, what one can hope is that such an investigation might give us a pointer, however faint,  to a future correct theory which incorporates fundamental length, quantum mechanics as well as theory of relativity. To that end, it appears that it would be useful to investigate other elementary quantum mechanics  problems \footnote{For a brief discussion of the free particle under the lattice assumption, see \cite{Bhatia}.}.  For instance, it would be interesting to bring out the $\lambda_0$-dependence of eigenvalue spectrum and eigenfunctions of a simple harmonic oscillator or hydrogen atom in our model and compare them with other approaches to consequences of fundamental length in QM  (see for example \cite{Maziashvili} and \cite{Akhoury}).  Equally interesting would be to investigate if $\lambda_0$-dependence of any physical quantity is measurable in any experiment in the foreseeable future.

In spite of the several investigations cited by Floratos and Leontaris \cite{Floratos}, the subject of lattice space-time has not been satisfactorily dealt with in the literature. In this paper, we have introduced the central difference as a means to discretize space-time, which must be distinguished from other differences introduced by Hildebrand \cite{Hildebrand}. This then allows us to model the 1+1 dimensional space-time as a lattice grid. In this manner we allow for the possible existence of a fundamental length, such as Planck length, resulting from combining quantum mechanics and classical general relativity.  We then ask the question: what is the consequence of this in familiar quantum mechanics, such as the problem in quantum mechanics of the particle in a box.

In the quantized space-time grid quantum mechanics of the particle in a box, we have obtained the eigenfunctions and eigenvalues. We have also obtained the uncertainty relation in the form of $\Delta x \Delta p$. We obtain corrections generally of the order of $\lambda_0^2$. We also make it evident that in the limit of continuum quantum mechanics, the usual forms are recovered. It must be noted that many results are made simpler by use of \emph{Mathematica} and we have verified that the results obtained by \emph{Mathematica} are reproduced by corresponding exact calculations.

\acknowledgments
 We thank Samir K. Bose for his comments on an earlier version of this paper. The first author is grateful to the faculty and staff of the Department of Mathematics, University of Maryland, College Park, MD for their support.

\end{document}